%% file: AINon-programmers.tex
\documentclass[acmsmall,nonacm=true,screen=true]{acmart}

\usepackage{multirow,tabularx}

\setcitestyle{round}

\setcopyright{acmcopyright}
\copyrightyear{2024}
\acmYear{2024}
\acmDOI{10.1145/123123123.12122}

\acmConference[arXiv]{arXiv '2x}{date}{arXiv}
\acmBooktitle{arXiv '2x, date, arXiv}
\acmISBN{978-1-4503-XXXX-X/2x/xx}

\usepackage{subfig}
\usepackage{tikz}
\usetikzlibrary{positioning, calc, shapes.geometric, shapes.symbols, arrows.meta, fit, backgrounds, intersections}
\usepackage{pgfplots}
\pgfplotsset{compat=1.16}
\usepackage{pgfplotstable}

\usepackage{pdflscape}

\begin{document}

\title{AI for non-programmers:}
\subtitle{Applied AI in the lectures for students without programming skills \textsuperscript{$\ddagger$}}

\author{Julius Schöning}
\email{j.schoening@hs-osnabrueck.de}
\affiliation{%
  \institution{Osnabrück University of Applied Sciences}
  \department{Faculty of Engineering and Computer Science}
  \city{Osnabrück}
  \country{Germany}
  \postcode{DE-49076}
}

\author{Tim Wawer}
\email{t.wawer@hs-osnabrueck.de}

\affiliation{%
  \institution{Osnabrück University of Applied Sciences}
  \department{Faculty of Management, Culture and Technology}
  \city{Lingen}
  \country{Germany}
  \postcode{DE-49809}
}

\author{Kai-Michael Griese}
\email{k-m.griese@hs-osnabrueck.de}

\affiliation{%
  \institution{Osnabrück University of Applied Sciences}
  \department{Faculty of Business Management and Social Sciences  }
  \city{Osnabrück}
  \country{Germany}
  \postcode{DE-49076}
}

\renewcommand{\shortauthors}{}
\renewcommand{\shorttitle}{}

\begin{abstract}
Applications such as \textit{ChatGPT} and \textit{WOMBO Dream} make it easy to inspire students without programming knowledge to use artificial intelligence (AI). Therefore, given the increasing importance of AI in all disciplines, innovative strategies are needed to educate students in AI without programming knowledge so that AI can be integrated into their study modules as a \textit{future skill}. This work presents a didactic planning script for applied AI. The didactic planning script is based on the AI application pipeline and links AI concepts with study-relevant topics. These linkages open up a new solution space and promote students' interest in and understanding of the potentials and risks of AI. An example lecture series for master students in energy management shows how AI can be seamlessly integrated into discipline-specific lectures. To this end, the planning script for applied AI is adapted to fit the study programs' topic. This specific teaching scenario enables students to solve a discipline-specific task step by step using the AI application pipeline. Thus, the application of the didactic planning script for applied AI shows the practical implementation of the theoretical concepts of AI. In addition, a checklist is presented that can be used to assess whether AI can be used in the discipline-specific lecture. AI as a \textit{future skill} must be learned by students based on use cases that are relevant to the course of studies. For this reason, AI education should fit seamlessly into various curricula, even if the students do not have a programming background due to their field of study.
\end{abstract}








\maketitle
$ $
{\footnotesize
\textsuperscript{$\ddagger$}  \textit{Translated from the German of:} Schöning, J.; Wawer, T.; Griese, K.-M. ``KI für Nicht-Programmierer*innen: Angewandte KI im Hörsaal für Studierende ohne Programmierkenntnisse'' in Zukunftsorientierte Lehre an der Hochschule Osnabrück --- Einblick in ausgewählte Lehrentwicklungsprojekte. Voneinander Lehren lernen (5) (2024). \href{https://nbn-resolving.org/urn:nbn:de:bsz:959-opus-52866}{urn:nbn:de:bsz:959-opus-52866}}
$ $ \\[5mm]

\section{Introduction}
As computer science becomes increasingly intertwined with everyday technologies, the importance of AI education continues to grow. The central importance of AI is recognized worldwide, and the development of AI skills in future generations is encouraged \citep{Koeller2022}. However, higher education in applicable AI poses a massive challenge, given the rapid spread of AI in all disciplines.

While \textit{science, technology, engineering, and mathematics} (STEM) teaching in schools focuses on the mathematical foundations of AI, e.g., probability theory, students are often unaware of the central importance of these foundations for practical AI. Against this background, ``AI for non-programmers'' is becoming increasingly important. Since dealing with AI is designated as a \textit{future skill} \citep{Suessenbach2021}, applied AI in the lecture hall should be an essential component of most courses of studies curricula. Given the need to integrate AI education into non-computer science areas, this work presents a general didactic planning script for applied AI. This planning script is a possible basis for answering the question of how students who do not have a programming background due to their field of study can be effectively trained in AI with study-relevant topics.

The presentation of AI in comprehensible contexts of the respective fields of study is intended to awaken a deeper understanding and greater interest among students. This approach is essential as it is the only way to bridge the gap between abstract mathematical formulas behind AI and real AI applications. Available self-study courses for AI usually start with the mathematical basics of a single-layer perceptron \citep{Allen2021} and, therefore, have no relation to the discipline-specific contexts of non-computer science courses of studies. Teaching AI in comprehensible discipline-specific contexts aligns well with the research findings of \citet{Allen2021}, who emphasize the importance of contextual learning for understanding complex issues. The step-by-step consideration of the AI application pipeline presented as a structure for a lecture series with three to four teaching units is in line with the findings of \citet{Kandlhofer2016}, who argue for curriculum structures that ensure inclusion and accessibility in AI education. The application of AI in various fields, from healthcare \citep{Haleem2019} to control engineering \citep{Schoening2023,Schoening2022} and agricultural engineering \citep{Schoening2023b,Schoning2021}, is presented as practical evidence of its universal relevance.

The sample lecture series for master students in energy management not only teaches theoretical AI concepts but also demonstrates their practical implementation so that students can experience the tangible effects of AI in real-life application cases. This lecture setting aligns with \citet{Kolb2014} experiential learning theories, which advocate learning by doing and reflecting. The example lecture series underlines the feasibility of teaching AI without extensive programming prerequisites, consistent with the findings of \citet{Martins2022}, who emphasize the importance of adaptable didactic approaches in AI education.

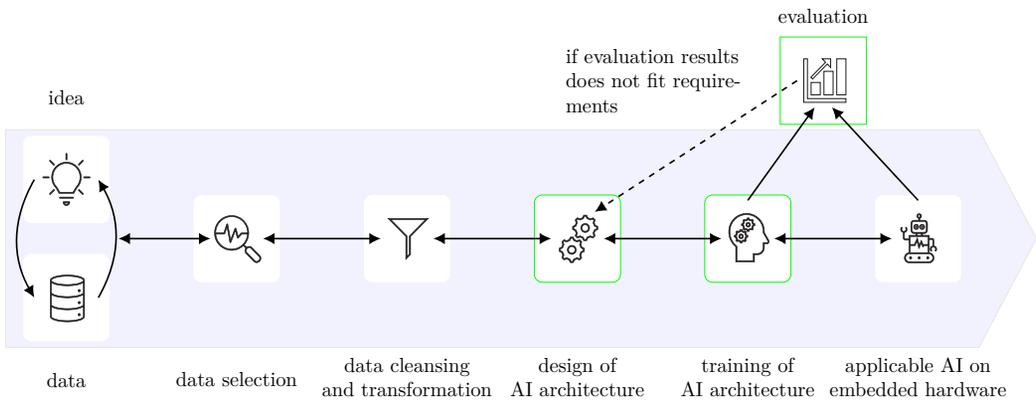
\begin{figure}[h]
  \centering
  \resizebox{\linewidth}{!}
    {
  \input{fig/pipelineTeaching}}
\caption{The AI application pipeline comprises six steps, with the steps in which programming is usually carried out outlined in green.}
\label{fig:pipeline}
\end{figure}

\section{AI Application Pipeline}
The development of AI can be mapped in a structured process with six steps \citep{Schoening2023a}, as shown in Fig.~\ref{fig:pipeline}. The first step is the iterative relationship between the data and application ideas. The elements lead to mutual refinement. Whether the idea or the data exists first depends on the application context. As soon as the application idea and the data have been determined, the second step is to select the data to solve the planned application. If AI is being used for the first time for a specific application idea, the question should be asked concerning the selected data as to whether a human could solve the tasks with the selected data. If a human can solve the tasks with the help of the selected data, the AI will quickly demonstrate valuable results, as the relevant patterns and information are then definitely available in the data. Following data selection, the third step is data cleansing. Data cleansing ensures that there are no imbalances in the data set. An imbalance can arise, e.g., during image data processing on which people are depicted. Here, the number of people of a particular gender, ethnicity, and age is decisive in each case and causes the imbalance. Ideally, the data cleansing in this example ensures that all genders are represented equally. The fourth step is to select the optimal \textit{AI model} for the task. The term \textit{AI architecture} is often used as a synonym for the \textit{AI model}.

Since new AI architectures are presented  are presented\footnote{
The categories of AI architectures can vary depending on how you look at them, but there are some general categories:
\begin{enumerate}
\item	Convolutional Neural Networks (CNN): These are often used for image and audio processing tasks and specialize in recognizing patterns in this data.
\item	Recurrent Neural Networks (RNNs): RNNs are designed for processing sequential data and are often used in text processing, language modeling and time series analysis.
\item	Transformer models: Transformer architectures, such as the Generative Pre-trained Transformer (GPT) model, are widely used in natural language processing and can be used as chatbots.
\item	Generative models: These models are designed to generate new data, such as Generative Adversarial Networks (GANs), which are used to create images and videos.
\item	Models of reinforcement learning: These models are used to solve decision problems by learning through interaction with an environment.
\item	Hybrid models: AI architectures that combine different techniques to handle more complex tasks.
\end{enumerate}
The list of categories is not exhaustive and will evolve over time as new architectures are constantly being developed.

} in the literature every day, there are overviews such as the \textit{Neural Network Zoo} \citep{Veen2019} and the \textit{Periodic Table of AI} \citep{Bitkom2023} to make it easier for users to select optimal \textit{AI models} for the intended application case. The most computer resource-intensive step is the training of the AI architecture, in which a large subset of the selected, cleansed, and adapted data is used for training, while a small subset of the data is retained for the evaluation of the AI. After successful training, the final step is to run the AI application in the application case using suitable software and hardware. For new application areas in particular, continuous evaluation and assessment should occur during training and use. If the evaluation results are inadequate to the intended application case, the \textit{AI model} must be adapted or changed and the training repeated, which is a resource-intensive cycle.

As illustrated in Fig.~\ref{fig:pipeline} with green frames, the AI application pipeline can be used without programming knowledge, except for the ``model selection'' and ``model training'' steps and the data-driven evaluation. The three steps, which otherwise regularly require programming knowledge, have been simplified for students using ready-made executable websites, so no prior computer science knowledge is required. The technical basis for these executable websites is the \textit{IPython Notebook} software package, which is available for Linux, Windows, and Apple computers.

\section{Didactic Planning Script for Applied AI in Discipline-Specific Contexts}
Three to four teaching units of 90 minutes each are required to integrate AI into the discipline-specific lecture or the corresponding course. Fig.~\ref{fig:script} shows the generalized didactic planning script for the teaching units on AI in discipline-specific contexts. AI can be seamlessly integrated into various curricula with the discipline-specific focus of the lecture or course.

\subsection{Teaching Unit 1: Application Idea and Data Set}
The first unit covers the first step of the AI pipeline. It is based on an application idea for which the corresponding data is sought or on data for which the appropriate application idea is sought. With regard to AI as a \textit{future skill}, this step is significant, as an understanding of AI's technical, legal and ethical limits must be acquired in conjunction with the application domain, i.e., the field of study, and an available or yet-to-be-created data set.

To promote students' motivation and interest, a quick introduction to AI should be achieved. Therefore, choosing a classification task as the application idea is advisable. Classification tasks are characterized by the fact that the data set is divided into different non-overlapping classes. An example of a classification task is the differentiation between the languages English, Klingon, and German. The data set for this task can consist of audio recordings, social media tweets, or essays containing the three languages. Another example of a classification task would be assigning images showing flowers to the botanically correct section: liverworts, mosses, hornworts, and vascular plants.

In the unlikely event that no classification task is found in the discipline-specific lecture, the continuation of a time series, the so-called time series prediction, can also be implemented with AI. In general, AI-based time series prediction is much more demanding to understand and is therefore provided as an optional fourth teaching unit in the didactic planning script.

\subsection{Teaching Unit 2: Data Analysis and Cleansing}
This unit covers two steps in the AI application pipeline. The data set is analyzed for errors and anomalies through visual inspection, i.e., a random check of the data set. Errors in data sets are omnipresent. These are caused, e.g., by the manual creation of data records, objects that cannot be clearly assigned to a specific class, or due to deliberate contamination of the data records by hackers. The data is selected based on the analysis. Incorrect data is removed from the data set and an equally \textit{weighted data set} is created. A weighted data set means all classes are equally represented in classification tasks. A weighted data set is non-discriminatory, as no class is favored. In a time series forecast data set, negative, positive, and neutral trends should occur with approximately equal frequency.
\pagebreak
\begin{landscape}
  \begin{figure}
  \resizebox{\linewidth}{!}{
  \footnotesize
  \begin{tabularx}{1.53\textheight}{|c|l|l|l|X|X|X|}
  \hline
  \textbf{No.}	& \textbf{Teaching Unit} &	\textbf{Example Lecture Title} &	\textbf{Learning Objectives} &	\textbf{Contents} &	\textbf{Methods} &	\textbf{Material} \\ \hline\hline
  \multirow{4}{*}{1} & \multirow{4}{*}{\begin{tabular}{l}application idea \\ and data set \end{tabular} } & \multirow{4}{*}{\begin{tabular}{l}
    Ventum Solution GmbH \\
     and Artificial Intelligence
   \end{tabular} } & \multirow{4}{*}{
   \begin{tabular}{l}- understanding the \\ possibilities of AI \\ - steps in the AI \\ application  pipeline \\ are known \\ - link between teaching \\ discipline-specific content \\ (application case) and \\ AI (data set) \end{tabular}
   } & Why is AI important today? & Case study & Case study  \\ \cline{5-7}
   & & & & Description of real AI application examples with discussion & Keynote speech + discussion & Slide set + Key questions \\ \cline{5-7}
   & & & & In which steps can AI be applied? & Lecture & AI application pipeline  \\ \cline{5-7}
   & & & & From the application idea to the data set & Small-group work & Sample data set, application idea  \\ \hline
   \multirow{3}{*}{2} & \multirow{3}{*}{\begin{tabular}{l}data analysis
 \\ and cleansing \end{tabular} } & \multirow{3}{*}{\begin{tabular}{l}
     Big data for  \\
      Ventum Solution GmbH
    \end{tabular} } & \multirow{3}{*}{
    \begin{tabular}{l}- evaluation of the quality \\ of data records\\
                    - selection of non-\\ discriminatory data \\
                    - starting the executable\\ web pages
 \end{tabular}
    } & Why is data the new gold? & Visual inspection & Data set with known errors \\ \cline{5-7}
    & & & & How fair are data sets? & Visual inspection & Data set with known imbalances \\ \cline{5-7}
    & & & & Executable web pages / \textit{IPython} & Practical work on the computer & Prefabricated \textit{IPython} web pages  \\ \hline
    \multirow{4}{*}{3} & \multirow{4}{*}{\begin{tabular}{l}AI for \\ classification \end{tabular} } & \multirow{4}{*}{\begin{tabular}{l}
      BirdImageScan - Wind  \\
       Power in Harmony With\\
       Nature Conservation
     \end{tabular} } & \multirow{4}{*}{
     \begin{tabular}{l}- overview of AI models\\ for classification \\
- discussions - 1) When is\\ AI ``good enough'' for a \\ particular application \\ case? 2) How expensive is \\ the use of AI?
 \end{tabular}
     } & How is AI trained and evaluated? & Practical work on the computer & Prefabricated \textit{IPython} web pages \\ \cline{5-7}
     & & & &  What AI models are there for classification & Self-learning task & Information material, link collection \\ \cline{5-7}
     & & & & Evaluation of AI models & Small-group discussion & Prefabricated \textit{IPython} web pages \\ \cline{5-7}
     & & & & Why can AI be harmful to the climate? & Classroom discussion & ``Warm'' and ``noisy'' lecture hall (due to PC fan)  \\  \hline
     \multirow{3}{*}{4} & \multirow{3}{*}{\begin{tabular}{l}AI for time series
   \\ forecasting \end{tabular} } & \multirow{3}{*}{\begin{tabular}{l}
       When is Electricity from  \\
        Wind and PV needed?
      \end{tabular} } & \multirow{3}{*}{
      \begin{tabular}{l}- overview of AI models \\ for time series forecasting \\
- discussion --- When is AI \\ ``good enough'' for this \\ specific application?
   \end{tabular}
      } & What does a time series data set look like? & Small-group work & Sample data set, application idea \\ \cline{5-7}
      & & & & What AI models are available for time series? & Self-learning task & Information material, collection of links \\ \cline{5-7}
      & & & & Which factors were not taken into account? & Small-group discussion & Prefabricated \textit{IPython} web pages  \\ \hline
  \end{tabularx}
  }
  \caption{Generalized didactic planning script for applied AI --- together with the discipline-specific content of the lecture or course, AI can be seamlessly integrated into various curricula.}
  \label{fig:script}
  \end{figure}
\end{landscape}
\pagebreak

If the AI is to be trained and evaluated on the students' laptops using the \textit{Bring Your Own Device} (BYOD) approach, it should be ensured that everyone can start the ready-made executable websites for this teaching unit. To this end, the \textit{IPython Notebook} software package should be installed using the Anaconda environment. In addition to the BYOD approach, using existing computer rooms in which \textit{IPython Notebooks} are installed is also possible. Operation within a cloud is also possible. Which option is used depends on the available hardware and the type of data set. When using a cloud service, it should be noted that uploading a non-public dataset may mean that the dataset becomes publicly accessible or that the cloud provider receives usage rights to this dataset.

\subsection{Teaching Unit 3: AI for Classification}
In this unit, it gets really loud, not due to the students, but to the fans of the computers and laptops, which have to ensure increased cooling capacity during the training of the AI models. An already functioning \textit{AI model} is selected for a quick start, and the students carry out the steps of model training and evaluation of this \textit{AI model} using the ready-made website. After the students have trained and evaluated their first AI model, they should get to know the step of model selection. To this end, the existing \textit{AI model} is to be adapted or even replaced by another model. This trial-and-error approach is popular with students, especially when the best \textit{AI model} is awarded a prize at the end of the unit. To ensure that the model selection is not just trial and error, explanations and links are provided on the pre-built executable webpage to help students select suitable AI models.

\subsection{Teaching Unit 4: AI for Time Series Forecasting}
This teaching unit follows the same pattern as the third teaching unit. It should be noted that, in contrast to the selection of the \textit{AI model} in the last teaching unit, the students only have minimal leeway to change the \textit{AI model} for time series forecasting to obtain the best possible evaluation results. The limited leeway results from the fact that the \textit{AI models} for time series forecasting have very complex structures, so without programming knowledge, there are only a few possibilities to select other \textit{AI models} or change the existing model. The few possibilities to customize the \textit{AI models} are explained with explanations and links, as described above, on the executable website.

\section{Example Lecture Series --- Ventum Solutions GmbH}
Based on the didactic planning script, see Fig.~\ref{fig:script}, a lecture series was developed for the ``energy management'' master's curriculum at Osnabrück University of Applied Sciences on the Lingen campus. A case study was created to link discipline-specific content and AI. In this case study, the fictitious German energy company Ventum Solutions GmbH wants to generate energy in an environmentally conscious way. Ventum Solutions GmbH and its subsidiaries want to
\begin{itemize}
  \item[a)] develop new areas for wind turbines and wind farms and
  \item[b)] feed green electricity into the script at the right time of day.
\end{itemize}

\noindent At the beginning of the first teaching unit, the students familiarize themselves with this case study and the challenges faced by Ventum Solutions GmbH.

\begin{figure}[h!]
  \centering
  \def\cnn at (#1,#2){
\draw[fill=black,opacity=0.2,draw=black] ($(#1)-(0.5*#2,0.5*#2)+2.5*(0.25*#2,0.25*#2)$) rectangle ++($(#2,#2)$);
\draw[fill=black,opacity=0.2,draw=black] ($(#1)-(0.5*#2,0.5*#2)+1.5*(0.25*#2,0.25*#2)$) rectangle ++($(#2,#2)$);
\draw[fill=black,opacity=0.2,draw=black] ($(#1)-(0.5*#2,0.5*#2)+0.5*(0.25*#2,0.25*#2)$) rectangle ++($(#2,#2)$);
\draw[fill=black,opacity=0.2,draw=black] ($(#1)-(0.5*#2,0.5*#2)-0.5*(0.25*#2,0.25*#2)$) rectangle ++($(#2,#2)$);
\draw[fill=black,opacity=0.2,draw=black] ($(#1)-(0.5*#2,0.5*#2)-1.5*(0.25*#2,0.25*#2)$) rectangle ++($(#2,#2)$);
\draw[fill=black,opacity=0.2,draw=black] ($(#1)-(0.5*#2,0.5*#2)-2.5*(0.25*#2,0.25*#2)$) rectangle ++($(#2,#2)$);
}

\def\cnnR at (#1,#2,#3){
\draw[fill=black,opacity=0.2,draw=black] ($(#1)-(0.5*#2,0.5*#3)+2.5*(0.25*#2,0.25*#2)$) rectangle ++($(#2,#3)$);
\draw[fill=black,opacity=0.2,draw=black] ($(#1)-(0.5*#2,0.5*#3)+1.5*(0.25*#2,0.25*#2)$) rectangle ++($(#2,#3)$);
\draw[fill=black,opacity=0.2,draw=black] ($(#1)-(0.5*#2,0.5*#3)+0.5*(0.25*#2,0.25*#2)$) rectangle ++($(#2,#3)$);
\draw[fill=black,opacity=0.2,draw=black] ($(#1)-(0.5*#2,0.5*#3)-0.5*(0.25*#2,0.25*#2)$) rectangle ++($(#2,#3)$);
\draw[fill=black,opacity=0.2,draw=black] ($(#1)-(0.5*#2,0.5*#3)-1.5*(0.25*#2,0.25*#2)$) rectangle ++($(#2,#3)$);
\draw[fill=black,opacity=0.2,draw=black] ($(#1)-(0.5*#2,0.5*#3)-2.5*(0.25*#2,0.25*#2)$) rectangle ++($(#2,#3)$);
}
\resizebox{\linewidth}{!}{
\begin{tikzpicture}
\node[inner sep=0pt,text width=5cm,draw=black,thick] (pos1){\includegraphics[width=\linewidth]{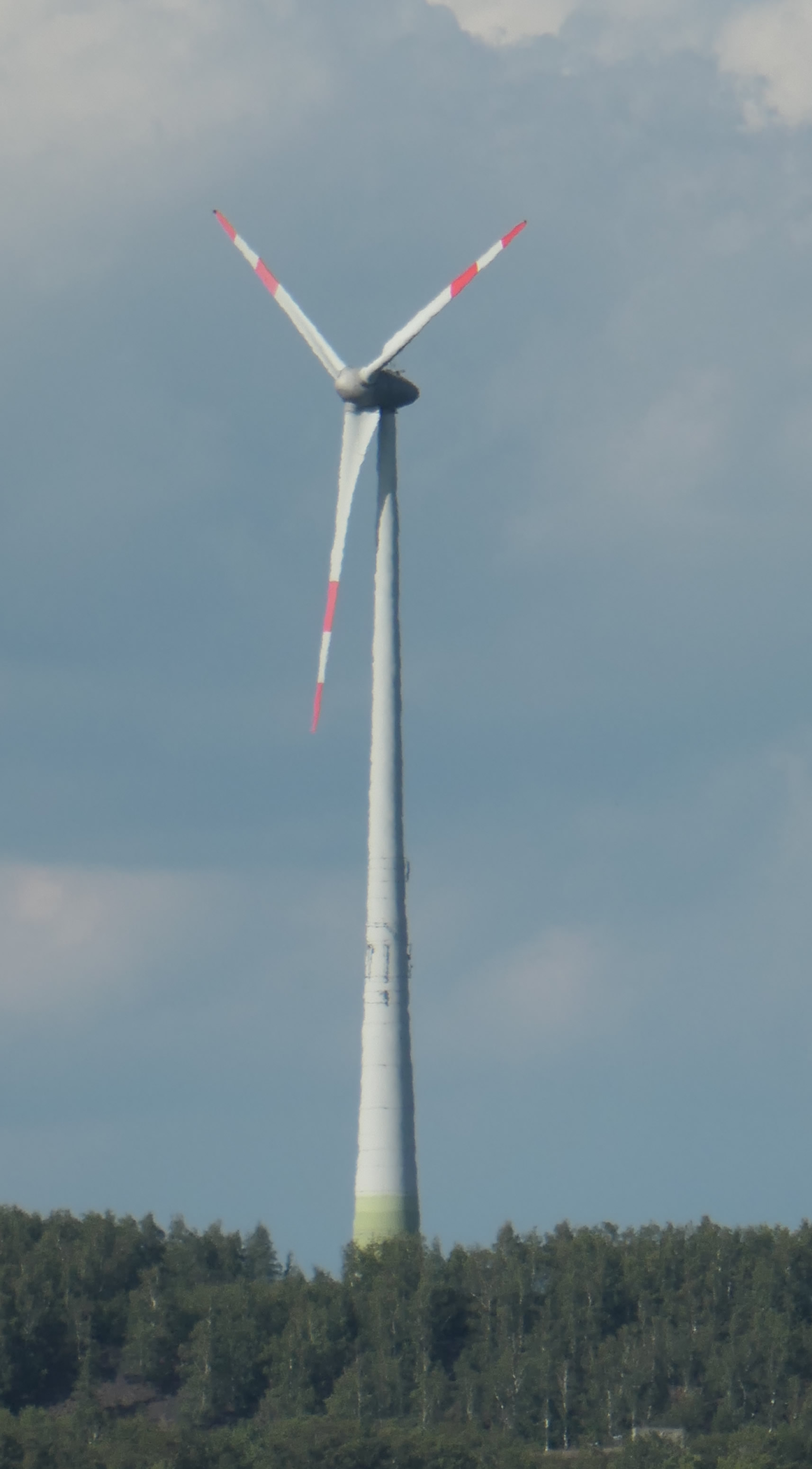}};


\node[above=0.05cm] at (pos1.north){\begin{tabular}{c}\textbf{Input}\end{tabular}};

 \coordinate (post1) at ($(pos1.center)+(5.0,0)$);
 \cnn at (post1,1.7);
 \node[above=0.55cm] at ($(post1)+(0.4,1.25)$){\begin{tabular}{c}Convolutional\\ Layer\end{tabular}};

  \coordinate (posC) at ($(post1.center)+(3,0)$);
 \cnn at (posC,1.3);
 \node[below=0.55cm] at ($(posC)-(0.6,0.95)$) {\begin{tabular}{c}Pooling\\ Layer\end{tabular}};

 \coordinate (posCC) at ($(posC.center)+(2.5,0)$);
 \cnn at (posCC,1.15);
 \node[above=0.55cm] at ($(posCC)+(0.3,0.65)$) {\begin{tabular}{c}Convolutional\\ Layer\end{tabular}};

\coordinate (posMergeMP) at ($(posCC.center)+(2.2,0)$);
\cnn at (posMergeMP,1.0);
 \node[below=0.55cm] at ($(posMergeMP)-(0.5,0.6)$) {\begin{tabular}{c}Pooling\\ Layer \end{tabular}};

\coordinate (posMergeDO) at ($(posMergeMP.center)+(1.8,0)$);
\cnn at (posMergeDO,0.8);
\node[above=0.55cm] at ($(posMergeDO)+(0.2,0.25)$) {\begin{tabular}{c}Convolutional\\ Layer \end{tabular}};

\coordinate (posFC) at ($(posMergeDO.center)+(1.8,-1.05)$);
\draw[fill=black,draw=black,opacity=0.5] ($(posFC)$) -- ($(posFC)+(0.5,0)$) -- ($(posFC)+(2,1.75)$) -- ($(posFC)+(1.5,1.75)$) -- ($(posFC)$);
\node [below=0.55cm] at ($(posFC)+(0.5,0.5)$){\begin{tabular}{c}Fully Connected\\Layer \end{tabular}};

\coordinate (posFC1) at ($(posFC.center)+(2.0,0.4)$);
\draw[fill=black,draw=black,opacity=0.5,scale=0.5] ($(posFC1)$) -- ($(posFC1)+(1.0,0)$) -- ($(posFC1)+(2.5,1.75)$) -- ($(posFC1)+(1.5,1.75)$) -- ($(posFC1)$);
\node [above=0.85cm] at ($(posFC1)+(0.75,0.5)$){\begin{tabular}{c}Fully Connected\\Layer \end{tabular}};

\coordinate (posOut) at ($(pos1.center)+(22,2*0.75)$);
\foreach \l [count=\i] in {Class Birds;, Class Miscellaneous;, Class Nothing;}{
\node[inner sep=0pt,below=\i*0.75cm,  anchor=west] (posOut\i) at (posOut){\begin{tabular}{l}{\l}\end{tabular}};
}
\node[above, anchor=west] at (posOut){\begin{tabular}{c} \textbf{Output} \end{tabular}};

\draw [->, bend angle=45, bend left, thick]  ($(posMergeDO.center)+(0.6,0.0)$) to node[midway, above] {\begin{tabular}{c} Flatten\end{tabular}} ($(posMergeDO.center)+(2.4,0.0)$);

\draw [->, thick]  ($(posFC1)+(1.4,0.8)$) to  ($(posOut1.west)+(0.0,0.0)$);
\draw [->, thick]  ($(posFC1)+(1.1,0.4)$) to  ($(posOut2.west)+(0.0,0.0)$);
\draw [->, thick]  ($(posFC1)+(0.8,0.0)$) to  ($(posOut3.west)+(0.0,0.0)$);

 \draw[decorate,decoration={brace,amplitude=10pt}, line width=1pt] let \p{A1}=(posMergeDO.south east), \p{S1}=(pos1.south west) in ($(\x{A1},\y{S1})+(0,-0.25)$) -- node[below=0.6cm] {feature identification}  ($(\x{S1},\y{S1})+(0,-0.25)$);
 \draw[decorate,decoration={brace,amplitude=10pt}, line width=1pt] let \p{A1}=(posOut.south west), \p{A2}=(posFC.south east), \p{S1}=(pos1.south west) in ($(\x{A1},\y{A2})+(0,-1.25)$) -- node[below=0.6cm] {\begin{tabular}{c}classification based on\\ the features\end{tabular}}  ($(\x{A2},\y{A2})+(-0.9,-1.25)$);

	\end{tikzpicture}
	}
 \caption{Example of the teaching unit --- simplified representation of an AI model for image classification.}\label{fig:anndesign}
\end{figure}
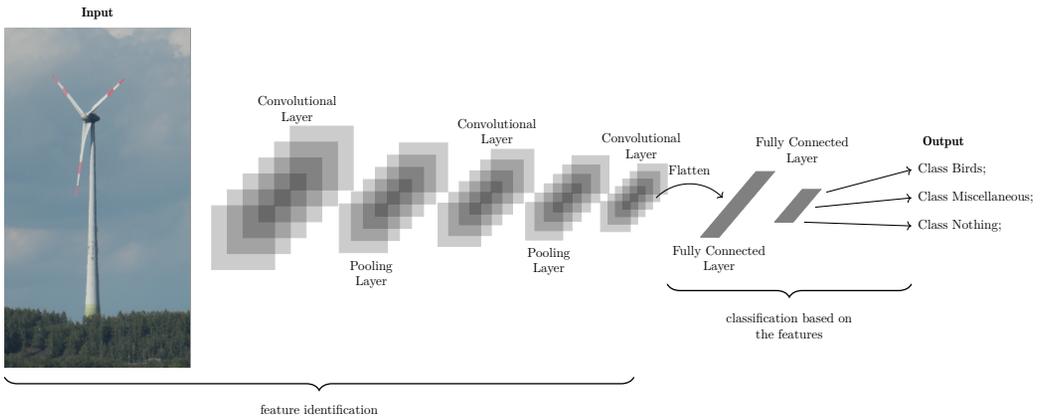

To enable Ventum Solutions GmbH to develop new areas for wind turbines and wind farms in nature reserves near airports, the students are to implement a new innovative method called BirdImageScan. As a classification task, BirdImageScan is designed to detect whether birds are in the vicinity of a wind turbine. The students have three classes of images: The ``Nothing'' class shows only the wind turbine, the ``Birds'' class shows one or more birds next to the wind turbine, and the ``Miscellaneous'' class shows airplanes, insects, and other objects next to the wind turbine. In the two teaching units, ``data analysis and cleansing'' and ``AI for classification'', the students develop an \textit{AI model} for image classification, as shown in Fig.~\ref{fig:anndesign}. If the students can prove that their developed \textit{AI model} can reliably recognize the class ``birds'', Ventum Solutions GmbH receives a permit to build wind turbines in nature reserves, as the wind turbines can switch off immediately as soon as a bird is detected thanks to BirdImageScan.

\begin{figure}[h]
  \includegraphics[width=.45\linewidth]{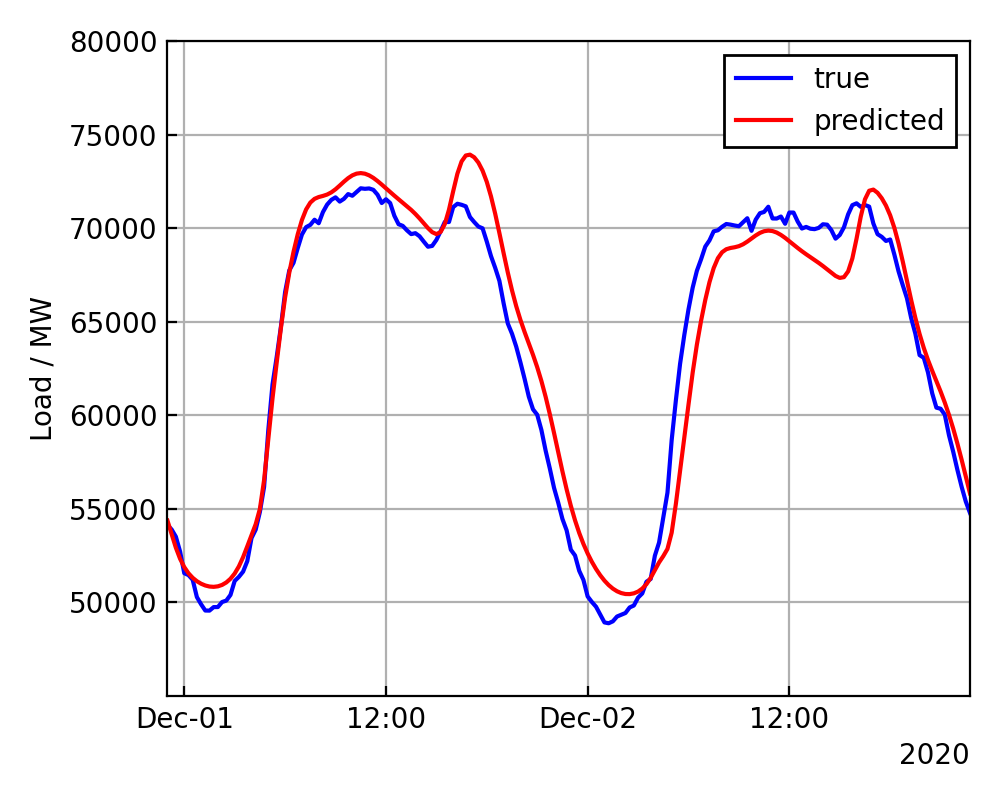}
  \caption{Example from teaching unit 4 --- result of the AI for time series forecasting. }
  \label{fig:lstm}
\end{figure}

Ventum Solutions GmbH would like to forecast electricity consumption to feed green energy into the grid in a targeted manner. The students are asked to create a time series forecast for the next 48 hours. In the ``AI for time series forecasting'' unit, the students optimize an \textit{AI model} with long-term and short-term memory, a so-called \textit{LSTM model}. As shown in Fig.~\ref{fig:lstm}, students can forecast consumption in the electricity grid using AI based on past consumption values with the help of their AI model. Critical reflection on the results shows that Ventum Solutions GmbH still needs to develop this \textit{AI model} further, as external influences such as weather, public holidays, and summer/winter time changes are not considered in the AI model.

\section{Checklist - AI in Discipline-Specific Courses}
Is AI suitable for my discipline-specific lecture or course? To answer this question, Fig.~\ref{fig:list} shows a short checklist to help lecturers determine whether AI can be integrated into the lecture. The integration of AI into the respective lectures should be coordinated with the entire curriculum, as the didactic planning script for applied AI shown in Fig.~\ref{fig:script} assumes that there has been no previous contact with AI and programming. If there have already been points of contact with AI and programming in other modules, the first two teaching units can be significantly condensed so that a total of two teaching units of 90 minutes each would be sufficient to achieve the learning objectives.

\begin{figure}[]
  \resizebox{\linewidth}{!}{
  \small

  \frame{
\begin{tabular}{lll}
\multicolumn{3}{c}{\textbf{Checklist: Applied AI in discipline-specific contexts}}                                                                                    \\
\multicolumn{3}{l}{Lecture / Course: \_\_\_\_\_\_\_\_\_\_\_\_\_\_\_\_\_\_\_\_\_\_\_\_}                                                                       \\
& &  \\
\multicolumn{2}{l}{Is there a classification task in the discipline-specific domain?}                                                          & $\square$  yes /$\square$  no \\[-1em]
\begin{tabular}{l} $ $\\
If yes:         \end{tabular}
            &                                                                                                          &             \\
                                    & Is this task understandable for the students?                                                            & $\square$  yes /$\square$  no \\
                                    & Does this task have at least two but no more than ten classes?                                           & $\square$  yes /$\square$  no \\
                                    & Are there at least 100 examples per class?                                                               & $\square$  yes /$\square$  no \\
                                    & Is there a freely accessible data set for this task?                                                     & $\square$  yes /$\square$  no \\
                                    & Is the available data set smaller than 4 gigabytes?                                                      & $\square$  yes /$\square$  no \\
                                    & & \\
\multicolumn{2}{l}{Is there a time series forecasting task in the discipline-specific domain?}                                                 & $\square$  yes /$\square$  no \\[-1em]
\begin{tabular}{l} $ $\\
If yes:         \end{tabular}                                &                                                                                                          &             \\
                                    & Is this task understandable for the students?                                                            & $\square$  yes /$\square$  no \\
                                    & Is there continuous time series data with a constant sampling rate?                                      & $\square$  yes /$\square$  no \\
                                    & Is there a freely accessible data set for this task?                                                     & $\square$  yes /$\square$  no \\
                                    & Is the time series in the data set at least twice as long as the prediction?                             & $\square$  yes /$\square$  no \\
                                    & Is the available data set smaller than 4 gigabytes?                                                      & $\square$ yes /$\square$  no \\
                                    & & \\
\multicolumn{2}{l}{\begin{tabular}{l}
  Do you have contact with a colleague with programming experience who\\ can support you in creating the executable web pages?
  \end{tabular}} & $\square$  yes /$\square$  no \\[-1em]
  &  & \\
\end{tabular}}
}
\caption{Checklist for whether practical AI can be used in the corresponding discipline-specific lecture or course.}
\label{fig:list}
\end{figure}

The guideline values of ten classes and 4 gigabytes are recommendations based on experience. They were developed taking into account the current performance of laptops to enable the execution of AI applications on students' laptops according to the BYOD approach. The question ``Are there at least 100 examples per class?'' should be understood to mean that there should be 100 examples for each class, e.g., 100 photos of roses, 100 photos of poinsettias, and 100 photos of bromeliads to serve as reference data.

\section{Conclusion and Outlook}
To put it bluntly, AI, like mathematics, is a problem-solving tool. Like mathematics, AI will also find its way into many disciplines, and its application will become increasingly commonplace. The presented didactic concept of how applied AI can be implemented in the lecture hall for students without programming knowledge is an excellent first guideline and creates initial points of contact between AI and the respective disciplines.

\begin{figure}[h]
  \includegraphics[width=\linewidth]{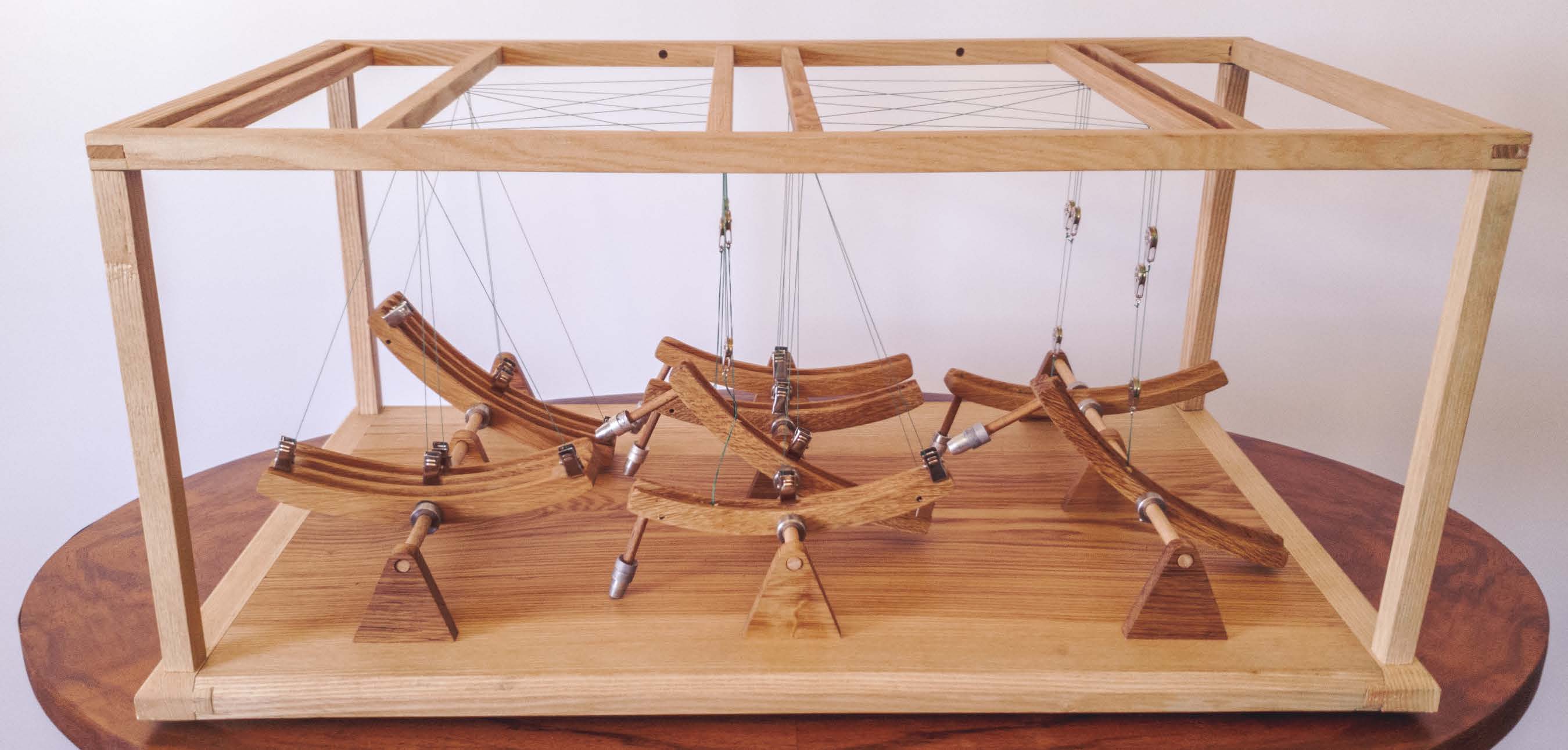}
  \caption{The mechanical neural network (MNN) is a physical model of an AI model and makes all components of AI literally tangible \citep{Schaffland2023}.}
  \label{fig:mnn}
\end{figure}

How could things continue after the potentials and risks of AI in the various application areas, as shown here in three to four teaching units, are understandable to the students? To what depth should students without programming knowledge, as non-computer scientists, learn to understand the background of AI? \citet{Schaffland2023} introduced the Mechanical Neural Network (MNN) to answer these questions. The MNN, as shown in Fig.~\ref{fig:mnn}, is a physical model of an AI model. The MNN makes all components of AI literally tangible and promotes deep AI understanding and interest. The MNN and the didactic value of a physical, tangible, and analog AI are currently being investigated, and learning concepts are being developed.

As with every newly designed teaching unit, the first prototype implementation of the sample lecture series ``Ventum Solutions GmbH'' still showed potential for improvement in some areas. In the second and third teaching units in particular, one or two tutors with programming knowledge should be on hand to provide support, as many detailed questions arise in the small student groups that can then be answered precisely. Conclusion: ``AI for non-programmers'' is possible and can be integrated into discipline-specific lectures and courses.

\section*{Funding Information}
Some of the work presented in this publication was part of the project ``Machine Learning zur Analyse großer Datenmengen am Beispiel der Energiewirtschaft''. Funding was provided as part of the ``Innovative Lehr- und Lernkonzepte: Innovation plus (2020/21)'' program of the German state Lower Saxony.




\bibliographystyle{ACM-Reference-Format}
\bibliography{ref}

\end{document}

%% file: fig/pipelineTeaching.tex
\begin{tikzpicture}
  \node(analysis)[draw=none, minimum width =1.5cm, minimum height =1.5cm, inner sep=0pt, rounded corners, fill=white]
  {\includegraphics[width=1cm]{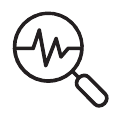}};
  \node(analysisText)[below = 0.9cm of analysis, rectangle, minimum height = 1.5cm] {data selection};
  \node(cleaning)[right = 1.45cm of analysis, draw=none, minimum width =1.5cm, minimum height =1.5cm, inner sep=0pt, rounded corners, fill=white]
  {\includegraphics[width=1cm]{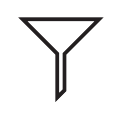}};
  \node(cleaningText)[below = 0.9cm of cleaning, rectangle, minimum height = 1.5cm] {\begin{tabular}{c}
  data cleansing \\ and transformation  \end{tabular}};
  \node(selection)[right = 1.45cm of cleaning, draw=green, minimum width =1.5cm, minimum height =1.5cm, inner sep=0pt,rounded corners, fill=white]
  {\includegraphics[width=1cm]{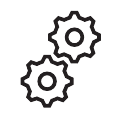}};
  \node(selctionText)[below = 0.9cm of selection, rectangle, minimum height = 1.5cm] {\begin{tabular}{c}
  design of \\ AI architecture  \end{tabular}};
  \node(training)[right = 1.45cm of selection, draw=green, minimum width =1.5cm, minimum height =1.5cm, inner sep=0pt,rounded corners, fill=white]
  {\includegraphics[width=1cm]{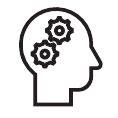}};
  \node(trainingText)[below = 0.9cm of training, rectangle, minimum height = 1.5cm] {\begin{tabular}{c}
  training of \\ AI architecture  \end{tabular}};
  \node(application)[right = 1.45cm of training, draw=none, minimum width =1.5cm, minimum height =1.5cm, inner sep=0pt,rounded corners, fill=white]
  {\includegraphics[width=1cm]{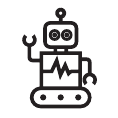}};
  \node(applicationText)[below = 0.9cm of application, rectangle, minimum height = 1.5cm] {\begin{tabular}{c}
  applicable AI on \\embedded hardware  \end{tabular}};
  \node(evaluation)[above right = 1.2cm and -0.2cm of training, draw=green, minimum width =1.5cm, minimum height =1.5cm, inner sep=0pt]
  {\includegraphics[width=1cm]{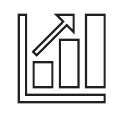}};
  \node(evaluationText)[above = -0.4cm of evaluation, rectangle, minimum height = 1.5cm] {evaluation};

  \node(idea)[above left  = -0.5cm and 1.45cm of analysis, draw=none, minimum width =1.5cm, minimum height =1.5cm, inner sep=0pt, rounded corners, fill=white]  {\includegraphics[width=1cm]{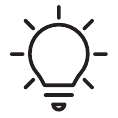}};
  \node(ideaText)[above= -0.1cm of idea, rectangle, minimum height = 1.5cm] {idea};

  \node(data)[below left  = -0.5cm and 1.45cm of analysis, draw=none, minimum width =1.5cm, minimum height =1.5cm, inner sep=0pt, rounded corners, fill=white]  {\includegraphics[width=1cm]{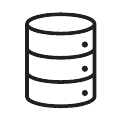}};
  \node(dataText)[below= -0.1cm of data, rectangle, minimum height = 1.5cm] {data};

   \coordinate (left0) at ($(idea.north west)+(-0.3,0.1)$);
   \coordinate (left1) at ($(data.south west)+(-0.3,-0.1)$);
   \coordinate (h0) at ($(application.east)+(0.3,0)$);
   \coordinate (h1) at ($(application.east)+(1.3,0)$);
   \coordinate (right0) at (left0-|h0);
   \coordinate (right1) at (left1-|h0);

   \begin{scope}[on background layer]
   \draw[black!10, fill=blue!5] (left0) -- (left1) -- (right1) -- (h1) -- (right0) -- cycle;
    \end{scope}

    \draw[-{Latex[length=2mm, width=2mm]}, thick, draw=black] ($(idea.west)+(0.2,0)$) to[bend right] ($(data.west)+(0.2,0)$);
    \draw[{Latex[length=2mm, width=2mm]}-, thick, draw=black] ($(idea.east)-(0.2,0)$) to[bend left] ($(data.east)-(0.2,0)$);
    \draw [{Latex[length=2mm, width=2mm]}-{Latex[length=2mm, width=2mm]}, thick, draw=black]($(analysis.west)+(-1.3,0)$)--($(analysis.west)+(0.3,0)$);
    \draw [{Latex[length=2mm, width=2mm]}-{Latex[length=2mm, width=2mm]}, thick, draw=black]($(analysis.east)+(-0.3,0)$)--($(cleaning.west)+(0.3,0)$);
    \draw [{Latex[length=2mm, width=2mm]}-{Latex[length=2mm, width=2mm]}, thick, draw=black]($(cleaning.east)+(-0.3,0)$)--($(selection.west)+(0.3,0)$);
    \draw [{Latex[length=2mm, width=2mm]}-{Latex[length=2mm, width=2mm]}, thick, draw=black]($(selection.east)+(-0.3,0)$)--($(training.west)+(0.3,0)$);
    \draw [{Latex[length=2mm, width=2mm]}-{Latex[length=2mm, width=2mm]}, thick, draw=black]($(training.east)+(-0.3,0)$)--($(application.west)+(0.3,0)$);

    \draw [-{Latex[length=2mm, width=2mm]}, thick, draw=black]($(application.north)+(0,-0.1)$)--($(evaluation.south)+(0.15,0.3)$);
    \draw [-{Latex[length=2mm, width=2mm]}, thick, draw=black]($(training.north)+(0,-0.1)$)--($(evaluation.south)+(-0.15,0.3)$);

    \draw [-{Latex[length=2mm, width=2mm]}, thick, dashed, draw=black]($(evaluation.west)+(0.3,0)$)--($(selection.north)+(0.3,-0.3)$);
    \node(evaluationFail)[left = 0cm of evaluation, rectangle, text width = 3.6cm] {if evaluation results does not fit requirements};
\end{tikzpicture}

%% file: AINon-programmers.bbl

\begin{thebibliography}{15}


\ifx \showCODEN    \undefined \def \showCODEN     #1{\unskip}     \fi
\ifx \showDOI      \undefined \def \showDOI       #1{#1}\fi
\ifx \showISBNx    \undefined \def \showISBNx     #1{\unskip}     \fi
\ifx \showISBNxiii \undefined \def \showISBNxiii  #1{\unskip}     \fi
\ifx \showISSN     \undefined \def \showISSN      #1{\unskip}     \fi
\ifx \showLCCN     \undefined \def \showLCCN      #1{\unskip}     \fi
\ifx \shownote     \undefined \def \shownote      #1{#1}          \fi
\ifx \showarticletitle \undefined \def \showarticletitle #1{#1}   \fi
\ifx \showURL      \undefined \def \showURL       {\relax}        \fi
\providecommand\bibfield[2]{#2}
\providecommand\bibinfo[2]{#2}
\providecommand\natexlab[1]{#1}
\providecommand\showeprint[2][]{arXiv:#2}

\bibitem[\protect\citeauthoryear{Allen, McGough, and Devlin}{Allen
  et~al\mbox{.}}{2021}]%
        {Allen2021}
\bibfield{author}{\bibinfo{person}{Becky Allen},
  \bibinfo{person}{Andrew~Stephen McGough}, {and} \bibinfo{person}{Marie
  Devlin}.} \bibinfo{year}{2021}\natexlab{}.
\newblock \showarticletitle{Toward a Framework for Teaching Artificial
  Intelligence to a Higher Education Audience}.
\newblock \bibinfo{journal}{\emph{{ACM} Transactions on Computing Education}}
  \bibinfo{volume}{22}, \bibinfo{number}{2} (\bibinfo{year}{2021}),
  \bibinfo{pages}{1--29}.
\newblock
\urldef\tempurl%
\url{https://doi.org/10.1145/3485062}
\showDOI{\tempurl}


\bibitem[\protect\citeauthoryear{{Bitkom e.V.}}{{Bitkom e.V.}}{2023}]%
        {Bitkom2023}
\bibfield{author}{\bibinfo{person}{{Bitkom e.V.}}}
  \bibinfo{year}{2023}\natexlab{}.
\newblock \bibinfo{booktitle}{\emph{Das Periodensystem der Künstlichen
  Intelligenz}}.
\newblock
\urldef\tempurl%
\url{https://www.periodensystem-ki.de/}
\showURL{%
\tempurl}


\bibitem[\protect\citeauthoryear{Haleem, Javaid, and Khan}{Haleem
  et~al\mbox{.}}{2019}]%
        {Haleem2019}
\bibfield{author}{\bibinfo{person}{Abid Haleem}, \bibinfo{person}{Mohd Javaid},
  {and} \bibinfo{person}{Ibrahim~Haleem Khan}.}
  \bibinfo{year}{2019}\natexlab{}.
\newblock \showarticletitle{Current status and applications of Artificial
  Intelligence ({AI}) in medical field: An overview}.
\newblock \bibinfo{journal}{\emph{Current Medicine Research and Practice}}
  \bibinfo{volume}{9}, \bibinfo{number}{6} (\bibinfo{year}{2019}),
  \bibinfo{pages}{231--237}.
\newblock
\urldef\tempurl%
\url{https://doi.org/10.1016/j.cmrp.2019.11.005}
\showDOI{\tempurl}


\bibitem[\protect\citeauthoryear{Kandlhofer, Steinbauer, Hirschmugl-Gaisch, and
  Huber}{Kandlhofer et~al\mbox{.}}{2016}]%
        {Kandlhofer2016}
\bibfield{author}{\bibinfo{person}{Martin Kandlhofer}, \bibinfo{person}{Gerald
  Steinbauer}, \bibinfo{person}{Sabine Hirschmugl-Gaisch}, {and}
  \bibinfo{person}{Petra Huber}.} \bibinfo{year}{2016}\natexlab{}.
\newblock \showarticletitle{Artificial intelligence and computer science in
  education: From kindergarten to university}. In
  \bibinfo{booktitle}{\emph{2016 {IEEE} Frontiers in Education Conference
  ({FIE})}}. \bibinfo{publisher}{{IEEE}}.
\newblock
\urldef\tempurl%
\url{https://doi.org/10.1109/fie.2016.7757570}
\showDOI{\tempurl}


\bibitem[\protect\citeauthoryear{Kolb}{Kolb}{2014}]%
        {Kolb2014}
\bibfield{author}{\bibinfo{person}{David~A Kolb}.}
  \bibinfo{year}{2014}\natexlab{}.
\newblock \bibinfo{booktitle}{\emph{Experiential learning: Experience as the
  source of learning and development}}.
\newblock \bibinfo{publisher}{FT press}.
\newblock


\bibitem[\protect\citeauthoryear{Köller, Thiel, van Ackeren, Anders,
  Becker-Mrotzek, Cress, Diehl, Kleickmann, Lütje-Klose, Prediger, Seeber,
  Ziegler, Kuper, Stanat, Maaz, and Lewalter}{Köller et~al\mbox{.}}{2022}]%
        {Koeller2022}
\bibfield{author}{\bibinfo{person}{Olaf Köller}, \bibinfo{person}{Felicitas
  Thiel}, \bibinfo{person}{Isabell van Ackeren}, \bibinfo{person}{Yvonne
  Anders}, \bibinfo{person}{Michael Becker-Mrotzek}, \bibinfo{person}{Ulrike
  Cress}, \bibinfo{person}{Claudia Diehl}, \bibinfo{person}{Thilo Kleickmann},
  \bibinfo{person}{Birgit Lütje-Klose}, \bibinfo{person}{Susanne Prediger},
  \bibinfo{person}{Susan Seeber}, \bibinfo{person}{Birgit Ziegler},
  \bibinfo{person}{Harm Kuper}, \bibinfo{person}{Petra Stanat},
  \bibinfo{person}{Kai Maaz}, {and} \bibinfo{person}{Doris Lewalter}.}
  \bibinfo{year}{2022}\natexlab{}.
\newblock \bibinfo{title}{Digitalisierung im Bildungssystem:
  Handlungsempfehlungen von der Kita bis zur Hochschule. Gutachten der
  Ständigen Wissenschaftlichen Kommission der Kultusministerkonferenz (SWK)}.
\newblock
\newblock
\urldef\tempurl%
\url{https://doi.org/10.25656/01:25273}
\showDOI{\tempurl}


\bibitem[\protect\citeauthoryear{Martins and Wangenheim}{Martins and
  Wangenheim}{2022}]%
        {Martins2022}
\bibfield{author}{\bibinfo{person}{Ramon~Mayor Martins} {and}
  \bibinfo{person}{Christiane Gresse~Von Wangenheim}.}
  \bibinfo{year}{2022}\natexlab{}.
\newblock \showarticletitle{Findings on Teaching Machine Learning in High
  School: A Ten - Year Systematic Literature Review}.
\newblock \bibinfo{journal}{\emph{Informatics in Education}}
  (\bibinfo{year}{2022}).
\newblock
\urldef\tempurl%
\url{https://doi.org/10.15388/infedu.2023.18}
\showDOI{\tempurl}


\bibitem[\protect\citeauthoryear{Schaffland and Schöning}{Schaffland and
  Schöning}{2023}]%
        {Schaffland2023}
\bibfield{author}{\bibinfo{person}{Axel Schaffland} {and}
  \bibinfo{person}{Julius Schöning}.} \bibinfo{year}{2023}\natexlab{}.
\newblock \showarticletitle{Mechanical Neural Network: Making AI Comprehensible
  for Everyone}. In \bibinfo{booktitle}{\emph{2023 IEEE 2nd German Education
  Conference (GECon)}}. \bibinfo{publisher}{{IEEE}}.
\newblock
\urldef\tempurl%
\url{https://doi.org/10.1109/gecon58119.2023.10295144}
\showDOI{\tempurl}


\bibitem[\protect\citeauthoryear{Schoning and Richter}{Schoning and
  Richter}{2021}]%
        {Schoning2021}
\bibfield{author}{\bibinfo{person}{Julius Schoning} {and}
  \bibinfo{person}{Mats~L. Richter}.} \bibinfo{year}{2021}\natexlab{}.
\newblock \showarticletitle{{AI}-Based Crop Rotation for Sustainable
  Agriculture Worldwide}. In \bibinfo{booktitle}{\emph{2021 {IEEE} Global
  Humanitarian Technology Conference ({GHTC})}}. \bibinfo{publisher}{{IEEE}}.
\newblock
\urldef\tempurl%
\url{https://doi.org/10.1109/ghtc53159.2021.9612460}
\showDOI{\tempurl}


\bibitem[\protect\citeauthoryear{Schöning and Pfisterer}{Schöning and
  Pfisterer}{2023}]%
        {Schoening2023}
\bibfield{author}{\bibinfo{person}{Julius Schöning} {and}
  \bibinfo{person}{Hans-Jürgen Pfisterer}.} \bibinfo{year}{2023}\natexlab{}.
\newblock \showarticletitle{Safe and Trustful AI for Closed-Loop Control
  Systems}.
\newblock \bibinfo{journal}{\emph{Electronics}} \bibinfo{volume}{12},
  \bibinfo{number}{16} (\bibinfo{date}{Aug.} \bibinfo{year}{2023}),
  \bibinfo{pages}{3489}.
\newblock
\showISSN{2079-9292}
\urldef\tempurl%
\url{https://doi.org/10.3390/electronics12163489}
\showDOI{\tempurl}


\bibitem[\protect\citeauthoryear{Schöning, Riechmann, and Pfisterer}{Schöning
  et~al\mbox{.}}{2022}]%
        {Schoening2022}
\bibfield{author}{\bibinfo{person}{Julius Schöning}, \bibinfo{person}{Adrian
  Riechmann}, {and} \bibinfo{person}{Hans-Jürgen Pfisterer}.}
  \bibinfo{year}{2022}\natexlab{}.
\newblock \showarticletitle{{AI} for Closed-Loop Control Systems}. In
  \bibinfo{booktitle}{\emph{2022 14th International Conference on Machine
  Learning and Computing ({ICMLC})}}. \bibinfo{publisher}{{ACM}}.
\newblock
\urldef\tempurl%
\url{https://doi.org/10.1145/3529836.3529952}
\showDOI{\tempurl}


\bibitem[\protect\citeauthoryear{Schöning, Wachter, and Trautz}{Schöning
  et~al\mbox{.}}{2023}]%
        {Schoening2023b}
\bibfield{author}{\bibinfo{person}{Julius Schöning}, \bibinfo{person}{Paul
  Wachter}, {and} \bibinfo{person}{Dieter Trautz}.}
  \bibinfo{year}{2023}\natexlab{}.
\newblock \showarticletitle{Crop rotation and management tools for every
  farmer?}
\newblock \bibinfo{journal}{\emph{Smart Agricultural Technology}}
  \bibinfo{volume}{3} (\bibinfo{date}{Feb.} \bibinfo{year}{2023}),
  \bibinfo{pages}{100086}.
\newblock
\showISSN{2772-3755}
\urldef\tempurl%
\url{https://doi.org/10.1016/j.atech.2022.100086}
\showDOI{\tempurl}


\bibitem[\protect\citeauthoryear{Schöning and Westerkamp}{Schöning and
  Westerkamp}{2023}]%
        {Schoening2023a}
\bibfield{author}{\bibinfo{person}{Julius Schöning} {and}
  \bibinfo{person}{Clemens Westerkamp}.} \bibinfo{year}{2023}\natexlab{}.
\newblock \showarticletitle{AI-in-the-Loop -- The impact of HMI in AI-based
  Application}. In \bibinfo{booktitle}{\emph{Embedded World Conference 2023}}.
  \bibinfo{publisher}{Weka Fachmedien}, \bibinfo{pages}{550--556}.
\newblock
\urldef\tempurl%
\url{https://doi.org/10.48550/ARXIV.2303.11508}
\showDOI{\tempurl}


\bibitem[\protect\citeauthoryear{Suessenbach, Winde, Klier, and
  Kirchherr}{Suessenbach et~al\mbox{.}}{2021}]%
        {Suessenbach2021}
\bibfield{author}{\bibinfo{person}{Felix Suessenbach}, \bibinfo{person}{Mathias
  Winde}, \bibinfo{person}{Julia Klier}, {and} \bibinfo{person}{Julian
  Kirchherr}.} \bibinfo{year}{2021}\natexlab{}.
\newblock \bibinfo{booktitle}{\emph{Diskussionspapier Nr. 3 -- {FUTURE SKILLS
  2021} -- 21 Kompetenzen für eine Welt im Wandel}}.
\newblock \bibinfo{type}{{T}echnical {R}eport}.
  \bibinfo{institution}{Stifterverband für die Deutsche Wissenschaft e.V.}
\newblock
\urldef\tempurl%
\url{https://www.stifterverband.org/download/file/fid/10547}
\showURL{%
\tempurl}


\bibitem[\protect\citeauthoryear{van Veen and Leijnen}{van Veen and
  Leijnen}{2023}]%
        {Veen2019}
\bibfield{author}{\bibinfo{person}{Fjodor van Veen} {and}
  \bibinfo{person}{Stefan Leijnen}.} \bibinfo{year}{2023}\natexlab{}.
\newblock \bibinfo{booktitle}{\emph{The Neural Network Zoo}}.
\newblock
\urldef\tempurl%
\url{https://www.asimovinstitute.org/neural-network-zoo/}
\showURL{%
\tempurl}


\end{thebibliography}
